\documentclass[aps,prl,superscriptaddress,showkeys,showpacs,twocolumn,longbibliography,nofootinbib]{revtex4-1}
\usepackage[utf8]{inputenc}
\usepackage{amsmath}
\usepackage{amsfonts}
\usepackage{amsthm}
\usepackage{amssymb}
\usepackage{float}
\usepackage{braket}
\usepackage{graphicx}
\usepackage{url}
\usepackage[colorlinks=true, pdfstartview=FitV, linkcolor=red, citecolor=blue, urlcolor=blue]{hyperref}
\usepackage{slashed}
\usepackage[normalem]{ulem}

\graphicspath{{./figures/}}

\newcommand{\be}{\begin{equation}}
\newcommand{\ee}{\end{equation}}
\newcommand{\bea}{\begin{eqnarray}}
\newcommand{\eea}{\end{eqnarray}}

\begin{document}
\title{Chiral catalysis of nuclear fusion in molecules}

\author{\mbox{Dmitri E. Kharzeev}}
\email[]{dmitri.kharzeev@stonybrook.edu}

\affiliation{Center for Nuclear Theory, Department of Physics and Astronomy, Stony Brook University, Stony Brook, New York 11794-3800, USA}
\affiliation{Department of Physics, Brookhaven National Laboratory, Upton, New York 11973-5000, USA}

\author{Jake Levitt}
\email[]{jakelevitt@cortexfusion.systems}
\affiliation{Cortex Fusion Systems, Inc., New York, NY 10128, USA}

\bibliographystyle{unsrt}

\begin{abstract}
At low energies, nuclear fusion is strongly affected by electron screening of the Coulomb repulsion among the fusing nuclei. 
It may thus be possible to catalyze nuclear fusion in molecules (\textit{i.e.,} to fuse specific nuclei \textit{in situ}) through quantum control of electron wave functions in intense laser fields. The circularly polarized (chiral) laser field can effectively squeeze the electron wave functions, greatly enhancing the screening in the spatial  region relevant for the fusion process. We estimate the corresponding fusion probabilities, and find that the proposed chiral catalysis of nuclear fusion in molecules may be observable, potentially with important practical applications.
\end{abstract}

\maketitle
\noindent \emph{Introduction} Harnessing the energy of nuclear fusion is one of the grand challenges of modern science. Under normal conditions, the Coulomb repulsion barrier keeps the nuclei apart and prevents them from fusing. Two main  ways to overcome this repulsion are vigorously pursued at present: magnetic confinement and inertial confinement, see \cite{ongena2016magnetic,betti2016inertial} for reviews. In both cases, the tunneling through the Coulomb barrier is enhanced by creating a plasma, and thus increasing the kinetic energy of the colliding nuclei. 

In this letter, we investigate whether the fusion can be made to occur \textit{in situ} (\textit{i.e.,} inside molecules), without first creating a plasma. We argue that this may be possible when the electrons in a molecule are controlled by a circularly polarized ultrafast  laser.  Our  idea stems from the fact that at low energies the fusion probability is exponentially sensitive to the screening of nuclear Coulomb potential. This screening in molecules is provided by the constituent electrons, and is normally not strong enough to allow  fusion -- otherwise, our world would not be stable! 

However, the electron wave functions in atoms and molecules can be efficiently controlled by lasers. In particular, in the field of a high-frequency, circularly polarized laser, the electrons can be made to move along toroidal orbits of radius that is significantly smaller than the Bohr radius of innermost electrons. As a result, the Coulomb screening should become greatly enhanced, and this may lead to observable fusion rates, potentially suitable for  a fusion reactor. The circularly polarized (chiral) nature of laser field is crucial for this mechanism of fusion, and we will thus refer to it as the {\it chiral catalysis}. 

Our proposal is related to two important earlier ideas. The first one is the muon-catalyzed fusion ($\mu \textrm{CF}$) proposed long time ago by Frank \cite{frank1947hypothetical}, Sakharov \cite{sakharov1948passive} and Zel'dovich \cite{zel1954reactions}. In $\mu$CF, the Coulomb screening is enhanced due to the presence of muons that are about 200 times heavier than electrons and thus move much closer to the nuclei than electrons. The $\mu$CF was discovered experimentally by Alvarez and collaborators in 1957 \cite{alvarez1957catalysis}. The practical use of $\mu$CF is however challenging due to the necessity to first produce a sufficient flux of muons, that moreover have a relatively short lifetime of about $2\ \mu$s; see \cite{breunlich1989muon,ponomarev1990muon} for reviews.

The second, much less known, idea was put forward by Saha, Markmann and Batista \cite{saha2012tunneling} who proposed that the probability of tunneling through the Coulomb barrier can be enhanced by modifying the nuclear wave packet by laser pulses. Unfortunately, the effect of laser radiation on nuclei at visible and even soft X-ray wavelengths and moderate intensities  is quite weak due to the smallness of the nuclear Compton wavelength (or, equivalently, due to mismatch of energies between the photons and nuclei). 

However, as we will discuss in detail below, lasers do have a strong effect on electron wave packets, and can be used to effectively turn electrons into heavier particles moving close to the nuclei and providing the needed screening of the Coulomb potential.

\vskip0.3cm

\noindent \emph{The effect of Coulomb screening on nuclear fusion} Nuclear fusion is governed by the competition between the short-range attractive nuclear force and the long-range Coulomb repulsion among the fusing nuclei. The importance of Coulomb repulsion at a given kinetic center-of-mass energy $E$ is quantified by the Sommerfeld parameter
\begin{equation}
    \eta (E) = \frac{Z_1 Z_2 e^2}{v(E)},
\end{equation}
where $v(E) = \sqrt{2E/\mu}$ is the relative velocity of the nuclei with charges $Z_1$ and $Z_2$ and masses $M_1$ and $M_2$; $\mu = M_1 M_2/(M_1 + M_2)$ is the reduced mass. For low energies (say, in KeV range), $v \ll 1$, and the Sommerfeld parameter is very large, $\eta \gg 1$ -- this means that Coulomb effects are extremely important.  

The fusion cross section can be written as \cite{burbidge1957synthesis}
\begin{equation}\label{sfact}
    \sigma(E) = \frac{S(E)}{E} \exp{[-2\pi \eta(E)]},
\end{equation}
where $S(E)$ is the ``astrophysical S-factor'' that accounts for nuclear interactions and (in the absence of near-threshold resonances) varies slowly with energy. The exponential in (\ref{sfact}) accounts for the suppression due to tunneling through the repulsive Coulomb potential
\begin{equation}\label{coul}
    U(r) = \frac{Z_1 Z_2 e^2}{r},
\end{equation}
evaluated in semiclassical WKB approximation, as originally derived by Gamow \cite{gamow1928quantentheorie}. The WKB Euclidean action is evaluated between the classical turning point $r_t = Z_1 Z_2 e^2/E$ determined by the condition $E=U(r_t)$ and  the nuclear-scale distance $r_c$ at which the compound nucleus is formed.

If the fusing nuclei are part of atoms or molecules, the Coulomb potential (\ref{coul}) is screened by electrons. The appropriate treatment of the screening effects depends on the initial energy $E$. At low energies, when the relative velocity of the nuclei $v(E)$ is much smaller than the Bohr velocity of bound electrons $v_e \sim Z\alpha c$ (with $\alpha = e^2/4\pi$), $v(E)\ll v_e$, the electron cloud can quickly adjust to the motion of the nuclei, and the adiabatic Born-Oppenheimer approximation is appropriate. 

In this adiabatic approximation, the effect of Coulomb screening can be properly accounted for by the screened potential, for which we will assume a simple Debye form
\begin{equation}\label{couls}
    U_s(r) = \frac{Z_1 Z_2 e^2}{r} \exp \left( -\frac{r}{r_D} \right),
\end{equation}
where $r_D$ is the screening radius. The effect of screening on the tunneling probability can be estimated following the approach originally proposed by Salpeter \cite{salpeter1954electron} and developed in \cite{assenbaum1987effects,hale1990nuclear,koonin1989calculated}; see \cite{aliotta2022screening} for a recent review. Namely, the height of the Coulomb barrier near the distance $r_c$ where the compound nucleus is formed can be estimated from 
\begin{equation}
    U_s(r) \simeq \frac{Z_1 Z_2 e^2}{r} \left(1-\frac{r}{r_D}\right) \equiv U(r) - U_0,
\end{equation}
where we have used the Taylor expansion near $r=0$, justified by $r_c \ll r_D$. The height of the Coulomb barrier thus gets reduced by the amount
\begin{equation}\label{shift}
    U_0 = \frac{Z_1 Z_2 e^2}{r_D} ,
\end{equation}
and the corresponding increase of the tunneling probability can be estimated by an effective increase of energy, $E \to E + U_0$. The ratio of penetration probabilities with and without the screening is thus given by
\begin{equation}\label{ratio}
    R_s(E) \equiv \frac{\exp{[-2\pi \eta(E + U_0)]}}{\exp{[-2\pi \eta(E)]}}.
\end{equation}

The quantity (\ref{shift}) can be interpreted as the difference between the energy of the compound atom and the sum of energies of atoms before fusion \cite{salpeter1954electron}. 
Numerically, for $d + ^2H$ we estimate from (\ref{shift}) $U_0 \simeq 24$ eV by putting $r_D$ equal to the Bohr radius $r_B = (m_e\ c\ \alpha)^{-1} \simeq 0.53\ {\AA} $ (we use the natural system of units with $\hbar = 1$ and $e^2 = \alpha \simeq 1/137$).
Extracting $U_0$ from the energy difference of compound atom and individual deuterium atom and hydrogen molecule (assuming an equally weighted 
combination of the lowest-energy gerade and ungerade wave functions for the electron) yields $U_0 \simeq 20.4$ eV \cite{shoppa1993one}, a reasonably close value.  

For heavier atoms, one can estimate (\ref{shift}) by putting the screening radius $r_D$ equal to the radius of the innermost electron orbit, $r_D \simeq r_B/Z_i$, where $r_B = (m_e c \alpha)^{-1} \simeq 0.53\ {\AA} $ is the Bohr radius (we use the natural system of units with $\hbar = 1$) \cite{assenbaum1987effects}. If $Z_2 \gg Z_1$, we need to consider the effect of electron screening only around the nucleus with charge $Z_2$ and can neglect the effect of screening around the other nucleus. For example, for the $p + ^{16}O \to ^{17}F$ fusion process with $Z_2=8$ and $Z_1=1$, we estimate $r_D \simeq r_B/Z_2$ and $U_0 \simeq e^2 Z_2^2/r_B \simeq 1.54$ keV.

The fusion cross sections are usually studied in experiment at energies $E \gg U_0$, when the factor (\ref{ratio}) can be approximately rewritten as
\begin{equation}
    R_s(E) \simeq \exp{\left[\pi \eta(E) \frac{U_0}{E}\right]}, \hskip0.3cm E \gg U_0.
\end{equation}
This factor leads to a dramatic $\sim \exp({\rm const}/E^{3/2})$ exponential enhancement of the fusion cross section at low energies confirmed by experiment, see e.g. \cite{Engstler:1988tfw}. 

In this letter we will mostly be  interested in the ultra-low energy limit of $E \ll U_0$. Performing Taylor expansion of the exponent in (\ref{ratio}) to leading order in $E/U_0$, we get
\begin{equation}\label{ratio1}
    R_s(E) \simeq \exp{\left[2 \pi \eta(E)\right]}\times \exp{\left[- 2 \pi Z_1 Z_2 e^2 \sqrt{\frac{\mu}{2 U_0}}\right]},
\end{equation}
\begin{equation*}
    E \ll U_0.
\end{equation*}
We can multiply (\ref{sfact}) by this factor to obtain the fusion cross section at ultra-low energies with the account of Coulomb screening effect:
\begin{equation}\label{sfact-sc}
    \sigma_{low}(E) \simeq \frac{S(E)}{E} \exp{\left[- 2 \pi Z_1 Z_2 e^2 \sqrt{\frac{\mu}{2 U_0}}\right]}.
\end{equation}
We can observe that the Coulomb screening leads to the replacement of the vanishingly small at $E \to 0$ Gamow factor $\exp({-{\rm const}/\sqrt{E}})$ by a small but finite, energy-independent ``screened penetration factor"
\begin{equation}\label{ratio-s}
    P^{low}_s \equiv \exp{\left[-2\pi Z_1 Z_2 e^2 \sqrt{\frac{\mu}{2 U_0}}\right]} .
\end{equation}

The value of this factor for the $p + {\rm ^{16}O} \to {\rm ^{17}F}$ fusion process that we considered above is $\exp{(-147.2)} \simeq 10^{-64}$. If instead we estimated the Coulomb suppression using the traditional unscreened Gamow formula $\exp{(-2\pi\eta(E))}$ at, say, $E= 10$ eV characteristic for intra-molecular motion, we would get a much smaller factor of $\exp{(-300)} \simeq 10^{-130}$. 
\vskip0.3cm

\noindent \emph{Fusion rates in molecules}  Let us check whether this difference can lead to any observable effects in molecules. 
Consider first the $p + d$ fusion in hydrogen deuteride molecule ${\rm HD}$. Using the screening radius $r_D = r_B$, we get $U_0 \simeq 24\ {\rm eV}$, corresponding to the screened penetration factor (\ref{ratio-s}) 
\begin{equation}
    P^{low}_s ({\rm HD}) \sim 10^{-102}.
\end{equation}
The stretch vibration frequency of the hydrogen deuteride molecule is about $\Omega({\rm HD}) \simeq 4,340\ {\rm cm^{-1}} \simeq 1\times 10^{14}\ {\rm s^{-1}}$ \cite{1976CaJ}. Assuming that the $p + d \to {\rm ^{3}He}$ fusion process occurs in water every time the Coulomb barrier is penetrated, we estimate the rate of fusion processes in hydrogen deuteride as
\begin{equation}
    \lambda_f ({\rm HD}) \simeq P^{low}_s ({\rm HD}) \times \Omega({\rm HD}) \simeq 10^{-88}\ {\rm s^{-1}}.
\end{equation}
A fusion rate per unit volume per unit time is obtained by multiplying this rate by the density of molecules. Assuming $n ({\rm HD}) \simeq 10^{24}\ {\rm cm^{-3}}$,
\begin{equation}
    f({\rm HD}) = n ({\rm HD})\times   \lambda_f ({\rm HD}) \simeq 10^{-64}\ {\rm cm^{-3} s^{-1}} ,
\end{equation}
which is clearly unobservable. Achieving the fusion rate $f({\rm HD}) \sim {\rm cm^{-3}\ s^{-1}}$ would require decreasing the screening radius by a factor of 18. This estimate is consistent with the results of a more sophisticated analysis based on the R-matrix method \cite{hale1990nuclear}.

Let us now perform an estimate for the $p + {\rm ^{16}O} \to {\rm ^{17}F}$ fusion in water.
The symmetric $O-H$ stretch frequency  in water molecule is $\Omega({\rm H_2 O}) \simeq 3,657\ {\rm cm^{-1}} \simeq 1\times 10^{14}\ {\rm s^{-1}}$ \cite{1976CaJPh..54..525D}. The rate of fusion processes in water can thus be estimated as 
\begin{equation}
    \lambda_f ({\rm H_2 O}) \simeq P^{low}_s ({\rm H_2 O}) \times \Omega({\rm H_2 O}) \simeq 10^{-50}\ {\rm s^{-1}}. 
\end{equation}
corresponding to the characteristic lifetime of a gaseous water molecule
\begin{equation}
    T_f({\rm H_2 O}) = \lambda_f^{-1} ({\rm H_2 O}) \simeq 3\times  10^{42}\ {\rm years}.
\end{equation}
A fusion rate per unit volume per unit time for the density of water $n ({\rm H_2 O}) \simeq 3\ 10^{22}\ {\rm cm^{-3}}$ is
\begin{equation}
    f({\rm H_2 O}) = n ({\rm H_2 O}) \lambda_f ({\rm H_2 O}) \simeq 3\times 10^{-28} {\rm cm^{-3} s^{-1}} . 
\end{equation}

This estimate explains the stability of water w.r.t. nuclear fusion. Is there a way to make the fusion rate observable? According to our result (\ref{ratio-s}), the key to enhancing the fusion rate is to increase $U_0$, \textit{i.e.,} to decrease the screening radius $r_D$ below its value $r_D({\rm H_2 O}) \simeq r_B/8 \simeq 0.07 \ {\AA}$. 
The rate $f({\rm H_2O}) \sim {\rm cm^{-3}\ s^{-1}}$ can be achieved by decreasing the screening radius by a factor of 5.
 
\vskip0.3cm

\noindent \emph{Chiral catalysis of fusion by a circularly polarized laser} Is there a way of squeezing the electron cloud by a factor of ten or a hundred? This may be possible by putting atoms in a laser field. 
Assuming that the laser frequency $\omega$ is large compared to the frequency of atomic transitions, the effective potential of the electron-nucleus interaction can be obtained by averaging over the fast oscillations of the laser field. 

For linearly polarized laser field, this leads to so-called ``atomic dichotomy" \cite{pont1990radiative}, when the spherical electron distribution splits into two lobes separated by the ``classical excursion parameter" 
\begin{equation}
   r_C=\frac{eE}{m_e \omega^2}, 
\end{equation}
where $E$ is the peak laser electric field. This parameter 
represents the distance  traveled by the electron under the influence of Lorentz force $m_e a = - eE$ in the oscillating electric field $E$, $a t^2/2 \sim a/\omega^2 \sim eE/(m_e \omega^2)$.

However, for our purposes a much more interesting case is the circularly polarized laser field. In this case, the electron distribution becomes ring-like \cite{gersten1976shift,choi2002intense}, with symmetry axis oriented along the propagation of light, and going through the nucleus. The radius of the ring is equal to $r_C$, which corresponds to the circular motion of the electron at angular frequency $\omega$ that is phase-locked to the electric field of the laser.
Indeed, for a circularly polarized laser field with frequency $\omega$ the classical equation of motion of the electron under the influence of the Lorentz force $F(t) = - e E(t)$ results in a circular motion in the transverse $(x,y)$ plane (assuming that the light propagates along the $z$ axis) with angular frequency $\omega$ and the radius $r_C$.

In the presence of the Coulomb potential $U = - Ze^2/r$, the effective electron-nucleus potential can be obtained by using Kapitza method \cite{kapitza1951dynamic}, where the slow motion of the electron in Coulomb field and fast motion in the laser field are separated. The resulting effective potential in $(1+1)$ dimensions is then \cite{nadezhdin1986highly,oks2000rydberg}
\begin{equation}\label{pondero}
    U_{eff} = U + \frac{<F^2>}{4 m_e \omega^2} + \frac{<F^2>}{4 m_e \omega^4} \frac{d^2U}{dx^2},
\end{equation}
where $<F^2>$ is the Lorentz force averaged over the fast motion. The second and third terms on the r.h.s. of (\ref{pondero}) represent the ponderomotive energy, i.e. the time-averaged energy of electron's harmonic motion. After a straightforward generalization to 3D case, the effective electron-nucleus potential (\ref{pondero})  becomes \cite{gersten1976shift,oks2000rydberg}
\begin{equation}\label{pot}
    U_{eff}(r) = - \frac{Z e^2}{r} + \frac{r_C^2}{r^3}\ P_2(\cos{\theta}),
\end{equation}
where $\theta$ is the angle between the radius vector and the direction in which the laser field propagates, and $P_2(\cos\theta)=\frac{1}{2}(3 \cos^2\theta -1)$ is the Legendre polynomial. At $\theta=\pi/2$, this potential exhibits strong attraction in addition to the Coulomb one, and leads to an electron cloud squeezed along the directions perpendicular to the direction of propagating light. This potential is mathematically equivalent to the gravitational potential experienced by a satellite orbiting a prolate planet \cite{oks2000rydberg}.
\vskip0.3cm

\noindent \emph{Required laser parameters}
What are the requirements on a laser needed to achieve $r_C$ that is smaller than the typical screening radius by an order of magnitude? 
For our treatment to hold, the laser frequency has to be larger than atomic transition frequencies, and the laser intensity has to be large enough to justify the semiclassical approximation. Combining these requirements with the condition for the smallness of $r_C$ in comparison to the radius of the innermost electron orbit in a molecule, we find that an ideal laser frequency is in the soft X-ray range.

It may however be possible to approach the desired regime even with the tabletop UV lasers.
For example, a UV Lyman-alpha laser with wavelength of 122 nm has the frequency of $\omega \simeq 2.4\ 10^{15}$ Hz. Laser electric field $E$ is related to the laser intensity $I$ by $I=\epsilon_0 c E^2/2$, where $\epsilon_0$ is the permittivity of free space. 
This leads to $I(W/cm^2) = 1.33\ 10^{-3} [E(V/cm)]^2$. Assuming $I = 10^{9}\  W/cm^2$, we get $E= 10^6\ V/cm$, and $r_C \sim 0.02 \ {\AA}$ which is about five times smaller than the screening radius in the water molecule. According to the estimates given above, this should lead to an observable fusion rate.

Unfortunately, it is not clear whether the semiclassical treatment that we used would hold at relatively low intensities. To answer this question, we plan to investigate the corresponding Floquet dynamics in the water molecule numerically. 

Let us note however that the available tabletop laser systems utilizing higher harmonic generation produce wavelengths down to 1 nm.  It can thus be possible to reduce the screening radius $r_C \sim \sqrt{I}/\omega^2$ using a suitable combination of intensity $I$ and frequency $\omega$ within the domain of applicability of our treatment.
\vskip0.3cm

\noindent \emph{Summary} Chiral control of electron screening in molecules by circularly polarized lasers may open a new pathway towards nuclear fusion. However further theoretical and experimental studies are necessary to establish whether the corresponding fusion rates are observable and suitable for practical applications. 
\vskip0.3cm
\emph{Intellectual property} The concepts, systems, apparati, and techniques articulated herein are protected by U.S. Patent No. 63/596,122.


\vspace{1cm}

\section*{Acknowledgement}
The work of D.K. was supported in part by the U.S. Department of Energy, Office of Science, Office of Nuclear Physics, Grants Nos. DE-FG88ER41450 and DE-SC0012704.

\bibliographystyle{utphys}
\bibliography{main}

\end{document}